\documentclass[aps,amsmath,amssymb, twocolumn,
nofootinbib]{revtex4}

\usepackage{graphicx}
\usepackage{color}
\usepackage{bbold}

\newcommand{\bit}{\begin{itemize}}
\newcommand{\eit}{\end{itemize}}

\newcommand{\f}{\frac}
\renewcommand{\>}{\right\rangle}
\newcommand{\<}{\left\langle}
\newcommand{\ba}{\begin{align}}
\newcommand{\ea}{\end{align}}
\newcommand{\be}{\begin{equation}}
\newcommand{\ee}{\end{equation}}
\newcommand{\bi}{\begin{itemize}}
\newcommand{\ei}{\end{itemize}}
\newcommand{\lf}{\left(}
\newcommand{\ri}{\right)}
\newcommand{\dd}{\mathrm{d}}

\begin{document}

\newcommand{\bra}[1]{\< #1 \right|}
\newcommand{\ket}[1]{\left| #1 \>}

\title{Length Distributions in Loop Soups}
\author{Adam Nahum and J. T. Chalker}
\affiliation{Theoretical Physics, Oxford University, 1 Keble Road, Oxford OX1 3NP, United Kingdom}
\author{P. Serna, M. Ortu\~no and A. M. Somoza}
\affiliation{Departamento de F\'isica -- CIOyN, Universidad de Murcia, Murcia 30.071, Spain}
\date{August 1, 2013}

\begin{abstract}
Statistical lattice ensembles of loops in three or more dimensions typically have phases in which
the longest loops fill a finite fraction of the system. In such phases it is natural to ask about the distribution of loop lengths. We show how to calculate moments of these distributions using $CP^{n-1}$ or $RP^{n-1}$ and $O(n)$ $\sigma$ models together with replica techniques. 
The resulting  
joint length distribution for macroscopic loops 
is Poisson-Dirichlet with a parameter $\theta$ fixed by the loop fugacity and by symmetries of the ensemble. We also discuss features of the length distribution for shorter loops, and use numerical simulations to test and illustrate our conclusions.\\\\
\noindent
\end{abstract}
\maketitle

{\it Introduction.} Statistical models for ensembles of loops arise in many areas of physics and
the probability distribution of loop lengths offers an important way of characterising them. 
One expects that, although properties of the shortest loops are model-specific, universal
features should emerge on longer scales. In this paper we present results for
loop length distributions, obtained using simple field-theoretic methods and tested via Monte Carlo simulation.
Considering systems in which some loops are extended, we find the joint length distribution
of macroscopic loops via a calculation of moments that is exact (though not rigorous) in the relevant limit, showing 
that it is Poisson-Dirichlet (PD), a distribution \cite{kingman} with many applications in 
statistics and probability theory.

Loop ensembles are generated in a wide variety of problems from statistical physics.
They are central objects in polymer theory \cite{degennes} and series expansions \cite{series}.
They appear in vertex models \cite{vertex}, in the Coulomb phase of frustrated magnets \cite{jaubert1,jaubert2,sondhi}, and in
studies of resonating valence bond wavefunctions \cite{rvb,alba}.
An important class consists of vortex lines in three-dimensional random fields. These are of interest in
settings ranging from cosmology \cite{cosmic strings} to liquid crystals \cite{liquid-crystals} and optics \cite{dennis}, and lattice formulations \cite{tricolour percolation} have long been used for precise numerical simulations. 
More broadly, there are parallels \cite{derrida} between problems involving the length distributions of cycles in random maps, the replica theory of spin glasses, and Levy flights.


Loops and cycles also appear naturally for many-body quantum systems 
viewed in imaginary time, when
particle trajectories make up strands that close either via creation and annihilation processes 
or under 
time-periodic boundary conditions \cite{feynman}.  A key application is to quantum Monte Carlo simulations, in which properties of a system are computed by sampling the loop 
ensemble \cite{QMC}.

Loop and cycle models have likewise attracted considerable attention in the mathematical physics and statistics literature, both as probabilistic representations of quantum spin systems \cite{aizenman}
and as statistical problems in their own right. In particular, 
for a mean field-like problem of cycles on the complete graph,
the distribution of cycle lengths
has been shown to be PD \cite{schramm}. Recently Ueltschi and collaborators \cite{ueltschi-review}
have made the striking conjecture, with support from simulations \cite{ueltschi}, that the same
form also applies for long loops in three-dimensional systems. In this paper we provide a field-theoretic derivation of this result.

As we discuss in the following, two  features are sufficient to fix the universal behaviour in these ensembles.  
One is whether or not the loops are directed. 
The second is the fugacity associated 
with loops in the ensemble.

Our starting point 
is a class of models for loops on lattices
that we have defined in detail elsewhere \cite{ortuno somoza chalker,prl,pre,undirected,prb}.
These models may be formulated for either
directed or undirected loops. They have configurations ${\cal C}$ consisting of 
completely-packed coverings of the lattice by loops,
in which each link is part of one and only one loop. 
Coverings carry a statistical weight that is a product of two factors.
One depends on the local arrangement of loops at nodes and 
a control parameter $p$. The other varies with the number of 
loops $|{\cal C}|$ in the configuration
as $n^{|{\cal C}|}$, giving loops a fugacity that can be generated by
summing over $n$ possible `colourings' of each loop. 
The models support phases of two types: one in which all loops are finite, and another in
which some loops are extended. Our concern here is only with
behaviour in extended phases, reached via suitable choices for $p$ and the lattice.

We have found \cite{prl,prb} that a continuum description of these lattice models is provided
by $\sigma$ models with the target space $CP^{n-1}$ or $RP^{n-1}$ depending on
whether loops are directed or undirected. Moreover, $RP^{n-1}$ and $O(n)$ models
are equivalent for our purposes, since the $Z_2$ gauge symmetry that distinguish them plays no role
in the extended phases we are concerned with. Here we show how the $\sigma$ model
formulation allows a calculation of all moments of the length distribution for extended
loops. In the extended phase the $\sigma$ model has long range order: the computations
reduce to a finite-dimensional averages over orientations of the order parameter, which can be
evaluated exactly. 

{\it Calculations.} An outline of these calculations is as follows. We discuss initially
directed loops, indicating changes needed for the undirected case later. 
The $\sigma$ model field $Q$ is an $n\times n$ traceless Hermitian matrix that can be parameterised
in terms of an $n$-component complex vector ${\bf z}$ of length ${\bf z}^\dagger {\bf z}=n$ as
\begin{equation}\label{Q}
Q^{\alpha \beta} = z^\alpha \overline{z}^\beta - \delta^{\alpha \beta}\,.
\end{equation} 
Configurational averages in dimension $d$ are computed using the weight $e^{-S}$ with
\be
S = \frac{1}{2g} \int {\rm d}^dx \,{\rm tr}(\nabla Q)^2\,.
\ee
The $n$ values of the indices $\alpha$ and $\beta$ arise from possible loop colours 
and elements of $Q$, used as observables,
select loop configurations from the ensemble.
For instance, $Q^{\alpha \beta}(x)$ for $\alpha\not=\beta$ is a two-leg operator that, at the point $x$, absorbs a strand of colour $\alpha$ and emits one of colour $\beta$. The
average $\langle Q^{12}(x_1) \ldots Q^{m1}(x_m)\rangle$ is hence proportional to the weight in the ensemble
for a loop to pass in order through the points $x_1 \ldots x_m$, with its colour taking the values $1$ to $m$ on successive segments, instead of being a free variable. 

We use these ingredients to calculate moments of the loop length distribution \cite{note}. Let $l_1,\,l_2 \ldots$ be the lengths of loops in a configuration, in decreasing order. Then 
\ba \notag
\int \dd^d x_1  \ldots \dd^d x_m & \< Q^{12}(x_1) \ldots Q^{m1}(x_m) \> \\ 
& \quad\qquad =\f{ A^m}{n\,(m-1)!}\bigg\langle {\sum}_i l_i^m \bigg\rangle,
\end{align}
where $A$ is the nonuniversal normalisation associated with the operator $Q$, the factor of 
$(m-1)!$ arises because the points $x_1 \ldots x_m$ appear in a prescribed order around the 
selected loop, and the factor of $n$ is because there is no sum on the colour of this loop.
In the ordered phase, which occurs at small $g$ for $d>2$, the dominant contribution to the above $m$-point function comes from the spatially constant part of $Q^{\alpha\beta}(x)$, which is 
reduced in magnitude by fluctuations compared to (\ref{Q}). We therefore set
\be\label{avge}
Q^{\alpha\beta}(x) \simeq B\, (z^\alpha \bar z^\beta-\delta^{\alpha\beta}),
\ee
where $B$ is determined by the strength of long-range order. The correlator is given by averaging  ${\bf z}$ over the sphere ${\bf z}^\dag {\bf z} = n$, representing possible directions of broken symmetry.
This yields
\ba
\notag
\< Q^{12}(x_1) \ldots Q^{m1}(x_m) \> 
& \simeq 
\, B^{m} \langle |z^1|^2  \ldots |z^m|^2 \rangle \\ 
& =  
\, B^{m} \f{n^{m} \Gamma(n)}{\Gamma(n+m)}.
\end{align}
Combining expressions, taking the system volume (the total length of all loops)
to be ${\cal L}$, and writing $nB/A= f $, we have
\be \label{one-loop moment}
{\cal L}^{-m}\< {\sum}_i l_i^m \> = f^{m} \f{n\Gamma(n) \Gamma(m)}{\Gamma(n+m)}\,.
\ee
Since we retained only the spatially uniform part of $Q^{\alpha \beta}(x)$ in Eq.~(\ref{avge}), this
result applies in the thermodynamic limit ${\cal L}\to  \infty$. 

To have full information on the joint probability distribution of loop lengths
we must calculate general moments of the form
\be \notag
\< {\sum}'_{i_1, \ldots, i_q} l_{i_1}^{m_1} \ldots l_{i_q}^{m_q} \>.
\ee
Here the prime on the sum indicates that $i_1,\ldots, i_q$ are \emph{distinct} loops. This average is easily related to the integral of the $(\sum_{k=1}^q m_k)$-point correlation function
\be \notag
\< \Gamma^{(1)}\lf x_1^{(1)}, \ldots, x_{m_1}^{(1)} \ri \ldots \Gamma^{(q)} \lf x_1^{(q)}, \ldots, x_{m_q}^{(q)} \ri \>,
\ee
in which $\Gamma^{(k)}$ forces the coordinates $x^{(k)}_1, \ldots x^{(k)}_{m_k}$ to lie on the same loop via a product of two-leg operators -- for example
$
\Gamma^{(1)}\lf x_1^{(1)}, \ldots, x_{m_1}^{(1)} \ri=
Q^{12}(x^{(1)}_1)\ldots Q^{m_1 1}(x^{(1)}_{m_1}).
$
Each $\Gamma^{(k)}$ uses a distinct set of spin indices, so that  $m_{\rm tot} \equiv \sum_{k=1}^q m_k$ distinct spin indices are used in total. The same reasoning as for the case $q=1$ gives
\begin{align} \label{full}\notag
{\cal L}^{-m_{\rm tot}}\big\langle 
{\sum}'_{i_1, \ldots, i_q}& l_{i_1}^{m_1}\ldots l_{i_q}^{m_q} 
\big\rangle \\
 =& f^{m_{\rm tot}}
 \f{ n^q  \Gamma(n)  \Gamma(m_1)\ldots \Gamma(m_q) }{\Gamma(n+m_{\rm tot} )}\,,
\end{align}
again in the thermodynamic limit.

These calculation require integer $m_{\rm tot} \leq n$.
It is useful to relax both the
upper limit on $m_{\rm tot}$ and the restriction to integers.  
The replica technique achieves the first, simply by evaluating moments for sufficiently large fugacity, 
then setting $n$ to the required value in final expressions. Supersymmetric $\sigma$ models \cite{prl,pre,prb,read saleur,candu et al} provide
an alternative route to the same conclusions. Separately, allowing non-integer $m$ leads to an
interpretation of the coefficient $f$: since loops with $l_i \sim {\cal L}$ dominate the left-hand side of Eq.~(\ref{one-loop moment}) for large $\cal L$ and $m>1$, by taking the limits ${\cal L}\to \infty$, then $m\to 1$, we find that $f$ is the fraction of links covered by extended loops. Moreover, by taking a similar limit in Eq.~(\ref{full}), with $q=2$ and $m_1,\,m_2\to 1$, and Eq.~(\ref{one-loop moment}) at $m=2$, one sees that this fraction has no fluctuations in the thermodynamic limit.

The expressions we have obtained for moments are those of the Poisson Dirichlet distribution, which
we now introduce. It may be defined \cite{kingman} as the limiting case of a Dirichlet distribution for 
$M$ variables when $M$ diverges. In turn, the Dirichlet distribution is a probability distribution on $M$ variables $y_i$ ($i=1,2, \ldots M$) satisfying $y_i\geq 0$ and the constraint $\sum_{i=1}^M y_i=1$, with the form
\begin{eqnarray}\label{dirichlet}
&& \frac{\Gamma(M\alpha)}{[\Gamma(\alpha)]^M}(y_1, y_2 \ldots y_M)^{\alpha-1}{\rm d}y_1 \ldots {\rm d}y_{M-1}\,.
\end{eqnarray}
The PD distribution is the limit $M\to\infty$ and $\alpha\to 0$ of the Dirichlet distribution with the
parameter $\theta=M\alpha$ held fixed, and one obtains from Eq.~(\ref{dirichlet}) in this limit
\begin{equation} \label{PDmom}
\big\langle 
{\sum}'_{i_1, \ldots, i_q} y_{i_1}^{m_1}\ldots y_{i_q}^{m_q} 
\big\rangle 
 =
 \f{ \theta^q  \Gamma(\theta)  \Gamma(m_1)\ldots \Gamma(m_q) }{\Gamma(\theta+m_{\rm tot} )}\,.
\end{equation}
Comparison of Eqns.~(\ref{full}) and (\ref{PDmom}) shows that: $(i)$ the normalised, ordered lengths $x_i=l_i/(f{\cal L})$ of extended loops are PD distributed, and $(ii)$ the PD parameter takes the value $\theta=n$ for directed loops with fugacity $n$. These are our central results.

The modifications required to treat undirected loops are straightforward: in this case the $\sigma$ model field $Q$ is parameterised as in Eq.~(\ref{Q}) but the vector ${\bf z}$ is constrained to be real.
Subsequent results hold, but with the replacement of $n$ by $n/2$.
In particular, the PD parameter for undirected loops is $\theta=n/2$.

To put these results in context it is useful to connect them with what is known about the 
statistics of loop lengths on shorter scales, by considering in full the 
distribution $P_{\rm link}(l)= {\cal L}^{-1} l \langle \sum_i \delta(l-l_i)\rangle$ 
for the length $l$ of a loop passing through a randomly selected link. 
Consider a system of linear size $L$, so that ${\cal L} \sim L^d$. The correlation length
$\xi$ for the system sets a scale beyond which loops are Brownian \cite{prl}.
At distances $l\ll \xi$ behaviour is either model-dependent (if $\xi$ is comparable to the lattice spacing) or critical (if $\xi$ is large, a regime we have discussed elsewhere \cite{prl}).
In the range $\xi \ll l \ll L^2$, where loops are insensitive to sample boundaries, their
Brownian character implies
\be\label{diffusive}
P_{\rm link} (l) = Cl^{-d/2}
\ee
with non-universal normalisation $C$.
For $L^2 \ll l \leq f{\cal L}$ the distribution crosses over to the PD form 
\be\label{PD1}
P_{\rm link}(l) = \theta {\cal L}^{-1} (1-l/(f{\cal L}))^{\theta - 1}
\ee
obtained from Eq.~(\ref{dirichlet}) by integrating over all but one $y_i$.
Note that Eqns.~(\ref{diffusive}) and (\ref{PD1}) match in order of magnitude at $l\sim L^2$, as they should. Note also that in order to see both regimes it is important that the boundary conditions allow only loops, not open strands.

{\it Simulations.} As a test and illustration of our results, we next present data from 
Monte Carlo simulations
of three-dimensional loop models described in detail in Ref.~\cite{prl}. 
By studying directed loops for integer fugacity $1\leq n \leq 8$ and undirected loops for $n=1$,
we are able to access values of the PD parameter $\theta=1/2$ and integer $1\leq \theta \leq 8$.
We use system sizes 
$40\leq L\leq 100$ and average over $10^5$ loop configurations.

Our results complement 
earlier simulations of a different model with fixed PD parameter $\theta = 1$ \cite{ueltschi}. 
It is interesting to note that there exist other numerical observations of the PD in
literature on loop problems, which have not previously been recognised in these 
terms: see e.g. Fig.~2 (right panel) of Ref.~\cite{jaubert1} 
for $\theta=1/2$, Fig 7 of Ref.~\cite{sondhi} for $\theta=1$ and Fig.~3 of Ref.~\cite{alba} for $\theta=2$.

\begin{figure}[!h] 
\centering
\includegraphics[height=2.7in]{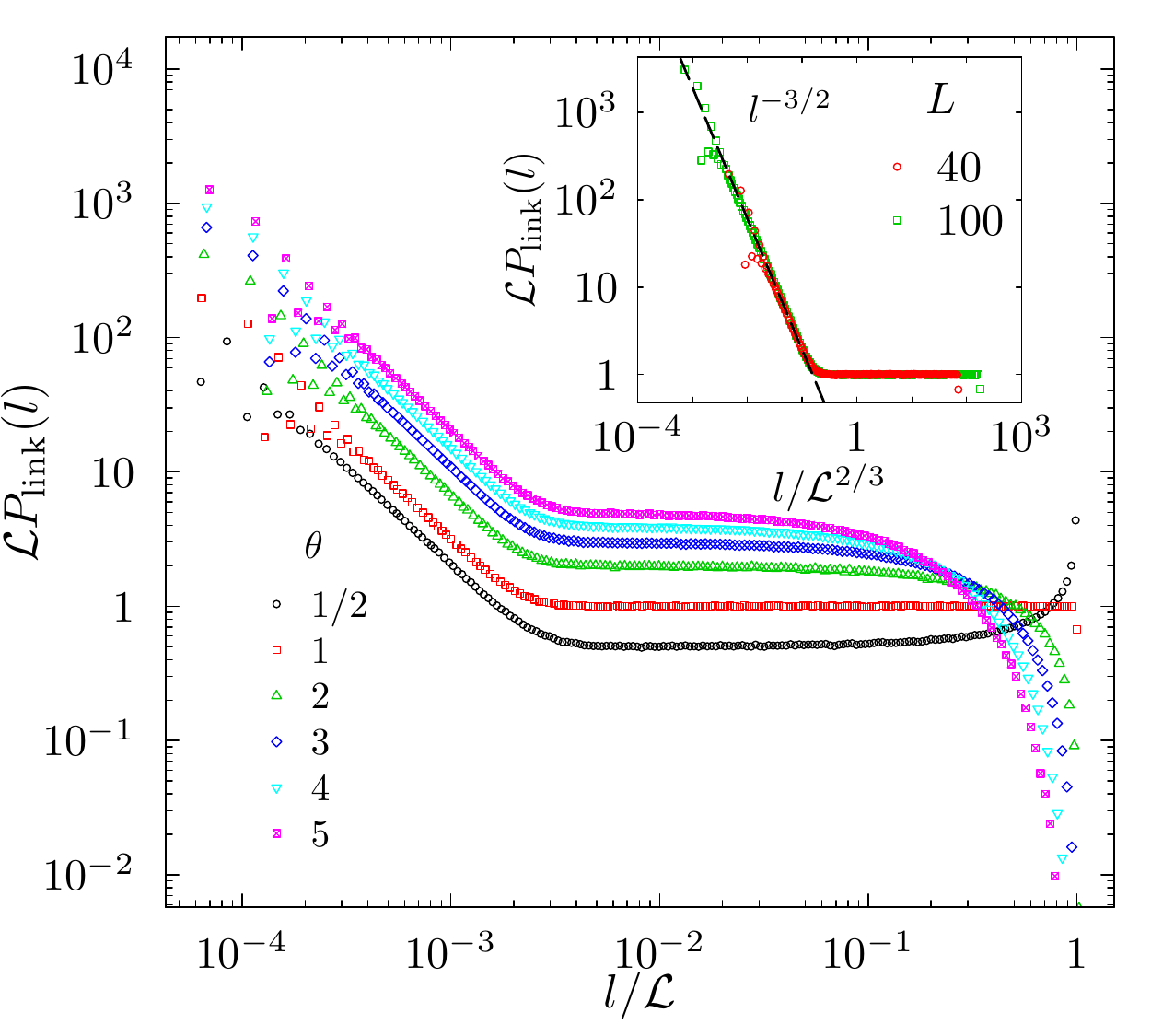} 
\caption{(Color online) ${\cal L}P_{\rm link}(l)$ vs. $l/{\cal L}$, on a double logarithmic scale for the indicated $\theta$ values. Inset: comparison of data for $L=40$ and $L=100$.  Dashed line has slope $-3/2$. Fluctuations at small $l$ arise from loops of length a few lattice spacings.}
\label{alldata}
\end{figure}

\begin{figure}[!h] 
\centering
\includegraphics[height=2.7in]{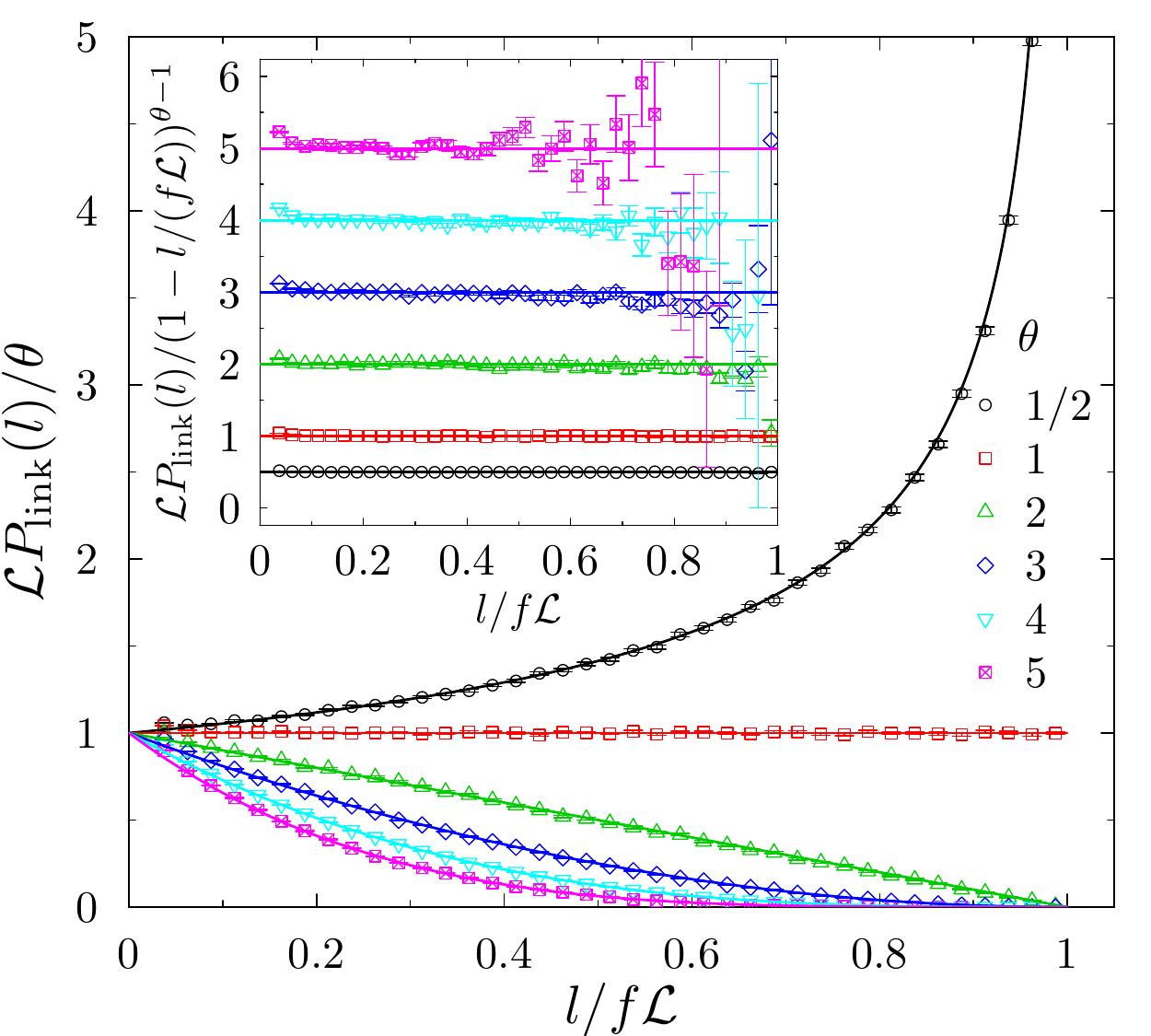} 
\caption{(Color online) Behaviour of loop length distribution in PD regime: ${\cal L} P_{\rm link}(l)/\theta$ vs. $l/f{\cal L}$, showing dependence on PD parameter $\theta$. Points: simulation data; curves from Eq.~(\ref{PD1}). Inset: the measured ratio ${\cal L}P_{\rm link}(l)/(1-l/(f{\cal L}))$ is close to its theoretical value $\theta$ except for $l\approx f{\cal L}$, where statistical errors are large.}
\label{distribution}
\end{figure}

We discuss first results for $P_{\rm link}(l)$. 
An overview of the calculated behaviour is given in Fig.~\ref{alldata}, where we plot ${\cal L}P_{\rm link}(l)$ versus $l/\mathcal{L}$ for size $L=40$. The two regimes of
Eqns.~(\ref{diffusive}) and (\ref{PD1}) are evident. Comparison of data from different system
sizes (see inset) demonstrates that the loop length for crossover between Brownian and PD forms
scales with $L^2$. In Fig.~\ref{distribution} we examine the PD regime  more closely, showing good agreement between data and theoretical expectations using the fraction $f$ of links on extended loops as the single fitting parameter. The dependence of the distribution on the PD parameter $\theta$ is striking.

\begin{figure}[!h] 
\centering
\includegraphics[height=2.7in]{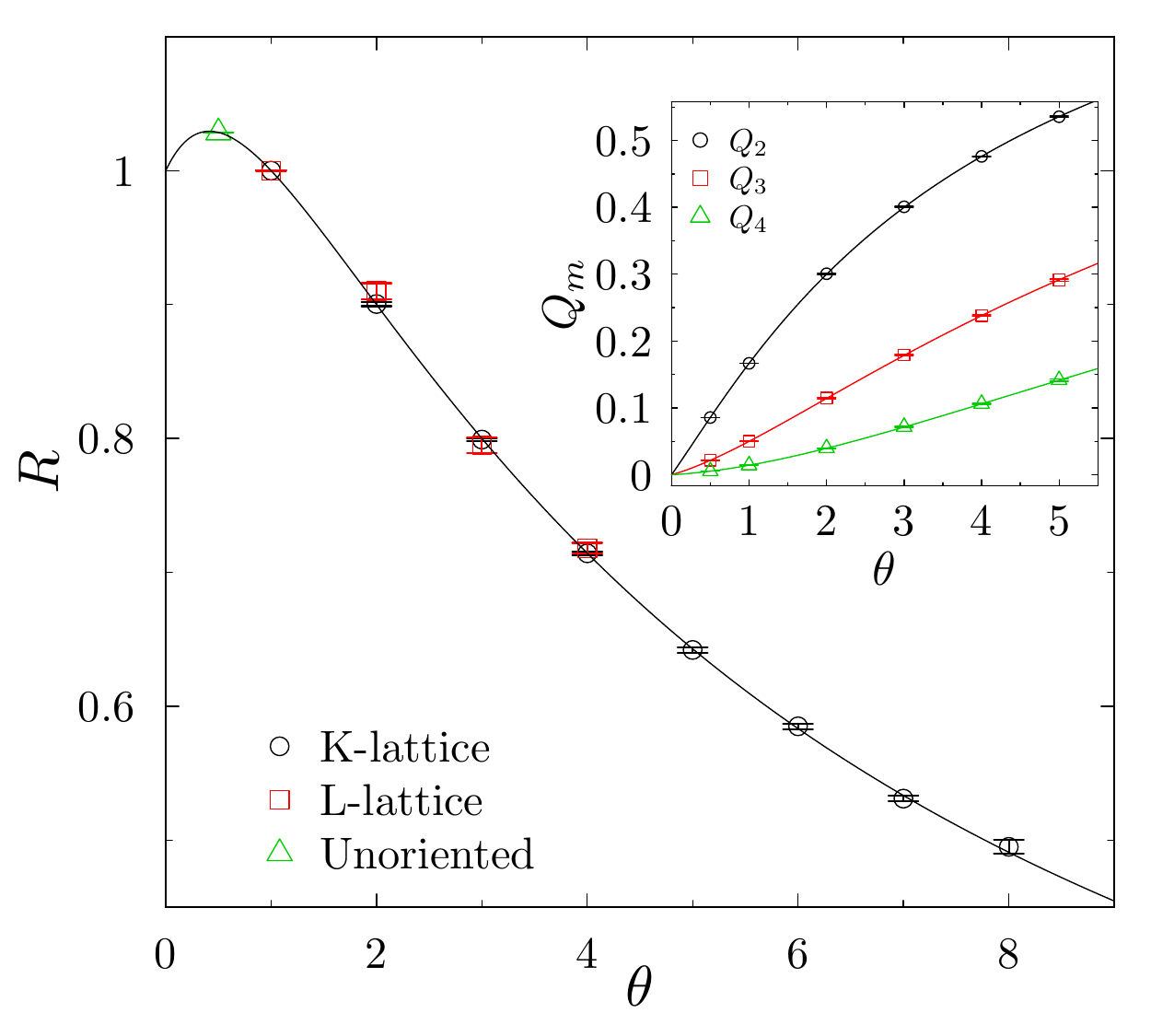} 
\caption{(Color online) Comparison of simulation data (points) with theoretical values (lines) for ratios of moments of loop lengths. In the main panel we plot the one-loop ratio $R$ and in the inset the two-loop ratios $Q_2$, $Q_3$ and $Q_4$. Universality is tested by the comparison of results for different
lattices (K, L and Unoriented), defined in Refs.~\cite{prl}, \cite{undirected} and \cite{prb}.}
\label{moments}
\end{figure}

As a further test, independent of any fitting parameter, we compute ratios of moments of the loop length distribution. We define
\begin{equation}
R\equiv  \frac{\langle \sum_{i} l_i^4 \rangle}{\langle \sum_{k} l_k^2 \rangle^2} \quad {\rm and} \quad
{\cal Q}_m \equiv \frac{\langle \sum_{i\not= j} l_i^m l_j^m \rangle}{\langle \sum_k l_k^m \rangle^2}\,.
\end{equation}
From Eq.~(\ref{full}) we expect 
\begin{equation}\label{ul}
R= \frac{6(\theta+1)}{(\theta+3)(\theta+2)}
\end{equation}
and
\begin{equation}
{\cal Q}_m =\frac{(m-1+\theta)(m-2+\theta) \ldots \theta}{(2m-1+\theta)(2m-2+\theta) \ldots (m+\theta)}\,.
\end{equation}
The simulation data shown in Fig.~\ref{moments} are in excellent agreement  with these expressions for $R$, $Q_2$, $Q_3$ and $Q_4$.


{\it Discussion.} It is remarkable that the joint probability distribution of lengths of macroscopic loops
should have a form that is both calculable and non-trivial. An appealing alternative perspective on this
result is provided by arguments that have been developed in the 
mathematical physics literature \cite{ueltschi-review}, which we now summarise. 

The starting point is to consider a stochastic process on loop configurations, of a type called split-merge, under which the distribution is stationary.  This process is in fact the one that we employ for Monte Carlo simulations \cite{prb}. At one time step we pick a node in the lattice at random and compare the colours of the two loop strands that pass through it. 
These strands are either two parts of a single loop that visits the node twice, or 
parts of two separate loops that each pass once through the node. If the colours are different we do not change the configuration, but if they are the same, then with a certain probability we break both strands and reconnect the resulting ends in a new pairing. In the case of
directed loops, there is only one way to form this new pairing
and the effect of the Monte Carlo move is either  (if both strands initially belong to a single loop) to 
split one loop into two, or (if the strands are initially from distinct loops) to merge two loops into one.
For undirected loops, since there are two ways to make new pairings, the Monte Carlo move always
merges two initial loops into one, but only splits a single loop into two on half of all attempts. 

The next stage is to derive consequences for the loop length distribution from the requirement that 
the rates at which loops split and merge should be equal. Let $\lambda_1$ and $\lambda_2$ be the lengths of
the split loops and $\lambda_1 + \lambda_2\equiv \lambda$ that of the merged loop. To make progress one needs to know 
the probabilities $\pi_1(\lambda)$ for the selected node to lie on one  loop, and
$\pi_2(\lambda_1,\lambda_2)$ for it to lie on two. If $\lambda_1$ and $\lambda_2$ are small, exact statements are impossible since the probabilities encode correlations of the loop ensemble. But when the loops are macroscopic ($\lambda_{1,2} \gg L^2$) it is 
plausible to conjecture \cite{ueltschi-review} that they pass through nodes in an uncorrelated way.
These probabilities are then simply related to lengths by the expressions
\begin{equation}
\pi_1(\lambda) = {\cal L}^{-2}\lambda \big\langle \sum_{i} \delta(\lambda - l_i) \big\rangle \equiv 
{\cal L}^{-1} P_{\rm link}(\lambda) \quad \label{pi1}
\end{equation}
and
\begin{equation}
\pi_2(\lambda_1,\lambda_2) = {\cal L}^{-2}\lambda_1 \lambda_2 \big\langle \sum_{i\not= j} \delta(\lambda_1 - l_i) \delta(\lambda_2 - l_j) \big\rangle\,. \label{pi2}
\end{equation}
For detailed balance we require
\begin{equation}
\theta \pi_1(\lambda_1+\lambda_2) =  \pi_2(\lambda_1,\lambda_2)\,,
\end{equation}
with $\theta=n$ for directed loops (since merging occurs only if both loops are the same colour) and
$\theta=n/2$ for undirected loops (when, in addition, the rate for splitting is halved).
Evaluating the right sides of Eqns.~(\ref{pi1}) and (\ref{pi2}) using the PD yields
$\pi_1(\lambda)= {\cal L}^{-2}\theta[1-\lambda
/(f{\cal L})]^{(\theta - 1)}$ and 
$\pi_2(\lambda_1,\lambda_2)= {\cal L}^{-2}\theta^2[1-
(\lambda_1+\lambda_2)
/(f{\cal L})]^{(\theta - 1)}$. 
Hence the PD is stationary under this split-merge process, provided long loops are sufficiently independent that the conjectured forms for $\pi_{1}(\lambda)$ and $\pi_{2}(\lambda_1,\lambda_2)$ are correct.

In summary, we have used the $\sigma$ model formulation of loop problems
and replica methods to establish a relation between the Poisson Dirichlet distribution 
of loop lengths and averages 
over the spaces $CP^{n-1} $ and $RP^{n-1}$ or $S^{n-1}$.

We thank Y. Fyodorov for a useful discussion. This work was supported in part by EPSRC Grant No.
EP/I032487/1 and by the MINECO and FEDER  Grants No. FIS2012-38206 and AP2009-0668.


\begin{thebibliography}{100}
\bibitem{note} We note that a similar idea for $m=2$ has been used in Ref.~\cite{ueltschi-proc} 
to relate the two-point correlation function of quantum spin models to their loop representations.
The extension to general $m$ necessitates the replica or supersymmetry methods that we introduce here.
\end{thebibliography}
\end{document}